\documentclass{IEEEcsmag}

\usepackage[colorlinks,urlcolor=blue,linkcolor=blue,citecolor=blue]{hyperref}
\usepackage{xcolor}
\usepackage{upmath}
\usepackage{listings}
\usepackage{balance}
\usepackage{booktabs}
\usepackage{multirow}
\usepackage[none]{hyphenat}

\jvol{XX}
\jnum{XX}
\paper{XX}
\jmonth{April}
\jname{XX}
\pubyear{2023}

\newcommand{\pset}{\vartheta}
\newcommand{\likelihood}{p(\vec{d}|\vec{\pset},H)}
\newcommand{\prior}{p(\vec{\pset}|H)}
\newcommand{\evidence}{p(\vec{d}|H)}
\newcommand{\posterior}{p(\vec{\pset}|\vec{d},H)}

\usepackage[utf8]{inputenc}

\begin{document}

\title{Reproducing the Results for NICER Observation of PSR~J0030+0451}

\author{\IEEEauthorblockN{
C. Afle\IEEEauthorrefmark{1},
P. R. Miles\IEEEauthorrefmark{1},
S. Ca\'ino-Lores\IEEEauthorrefmark{2}, 
C.~D.~Capano\IEEEauthorrefmark{3}\IEEEauthorrefmark{4},
I. Tews\IEEEauthorrefmark{5},
K. Vahi\IEEEauthorrefmark{6},
E.~Deelman\IEEEauthorrefmark{6},
M.~Taufer\IEEEauthorrefmark{2}}, and
D.~A.~Brown\IEEEauthorrefmark{1}
\IEEEauthorblockA{Syracuse U.\IEEEauthorrefmark{1}, 
U. Tennessee, Knoxville\IEEEauthorrefmark{2}, 
Max Planck Institute for Gravitational Physics\IEEEauthorrefmark{3}, 
U. Massachusetts Dartmouth\IEEEauthorrefmark{4}, 
Theoretical Division, Los Alamos National Laboratory\IEEEauthorrefmark{5}, 
and U. Southern California\IEEEauthorrefmark{6}\\
Emails: dabrown@syr.edu}}

\date{April 2023}

\begin{abstract}
NASA's Neutron Star Interior Composition Explorer (NICER) observed X-ray emission from the pulsar PSR J0030+0451 in 2018. Riley et al. reported Bayesian parameter measurements of the mass and the star's radius using pulse-profile modeling of the X-ray data. This paper reproduces their result using the open-source software X-PSI and publicly available data within expected statistical errors. We note the challenges we faced in reproducing the results and demonstrate that the analysis can be reproduced and reused in future works by changing the prior distribution for the radius and the sampler configuration. We find no significant change in the measurement of the mass and radius, demonstrating that the original result is robust to these changes. Finally, we provide a containerized working environment that facilitates third-party reproduction of the measurements of mass and radius of PSR J0030+0451 using the NICER observations. 
\end{abstract}
\maketitle


\section{INTRODUCTION}\label{sec:introduction}
Reproducibility of research---the ability to arrive at a consistent result given the same raw data and original analysis method---is a critical element of scientific discovery. Reproducibility provides the necessary level of trust in the published results and enables researchers to build upon that work. 
Since more and more scientific studies are using computation as a tool, reproducibility challenges arise from the computational point of view---- especially the availability of data, software, the needed execution environment, and tools, as well as documentation used in the original analysis~\cite{NAP25303}. 

NICER is a payload onboard the International Space Station and the X-ray Timing Instrument (XTI) is dedicated to observing X-rays from galactic pulsars~\cite{2016SPIE.9905E..1HG}. Based on NASA's open science and open data policy, the data observed by NICER is released to the public to advance scientific research. One of NICER's aims is to measure the masses and radii of neutron stars. These measurements constrain the neutron-star equation of state, the relation between the pressure and density of the neutron star. Measuring this equation of state requires a computationally intensive analysis of the NICER data. To fully understand and leverage the results of the equation of state analyses, the astrophysics community needs to be able to reproduce and modify the original results so that they can (i) check the robustness of the original result, (ii) build new analyses using the original result, or (iii) extend the original analysis to address new and different questions.

To measure the equation of state using PSR J0030+0451, Riley {\em et al.} developed the analysis software \textit{X-PSI} \footnote{\url{https://github.com/xpsi-group/xpsi.git}} (X-Ray Pulse Simulation and Inference)~\cite{xpsi}. \textit{X-PSI} includes a Bayesian analysis framework to measure the pulsar's mass and equatorial radius (hence infer the equation of state) using the observed NICER data. We explore whether the analysis of the pulsar PSR J0030+0451 by Riley \textit{et al.}~\cite{Riley:2019yda} can be reproduced and modified to test the robustness of the result. 

Miller \textit{et al.}~\cite{Miller:2019cac} has the same NICER observations of PSR J0030+0451 to produce an independent analysis using different software, models, and methods to measure the mass and radius of the pulsar. They arrived at measurements of mass and radius of PSR~J0030+0451 that were slightly different from, but consistent with, the results of Riley \textit{et al.} This is an example of \textit{replicability} of research: using the same data but different methods to arrive at a consistent result. The conclusion can be drawn that the results are replicable; an analysis of the data leads to a consistent mass and radius result for the pulsar. However, it does not verify that an external entity could use the existing software stack created by Riley \textit{et al.} to achieve the same result, nor does it demonstrate that another group could modify or extend this analysis.

Unlike our previous work on reproducing the detection of GW150914 by LIGO~\cite{Brown:2020emp}, none of the authors of this reproducibility effort were involved in the original analysis. This work is entirely based on the papers, data, software, and documentation provided to the public by the authors of the original study by Riley \textit{et al.} First, we reproduce the results in Figure 19 of Riley \textit{et al.}, which shows the measurement of the mass and the radius of the target pulsar obtained from the analysis. We note the lessons learned and challenges we faced during the reproducibility process, as was done in our previous works where we reproduced the images of the M87 black hole published by the EHT collaboration~\cite{Patel:2022acr, Ketron2021} and reproduced the detection of GW150914~\cite{Brown:2020emp}. We discuss the challenges encountered while acquiring the input data, installing and using the software (including setting up the required dependencies and environment), writing configuration files, job submission scripts, and post-processing the job output. Ultimately, we were successfully able to reproduce the measurements done in the original analysis.

Going beyond our previous work, after reproducing the original analysis, we demonstrate that Riley~\textit{et al.} provide sufficient information to allow a third party to modify the analysis in the new work. We use this functionality to test the robustness of the methods to the prior probability distributions chosen for the Bayesian analysis. Specifically, we expand the previous space on the pulsar radius from 16 km to 25 km and change the sampler configuration for the Bayesian analysis. 
We find that changing the upper limit of the prior does not change the posterior distribution statistically significantly, demonstrating the result's robustness to the prior radius choice. We increase the number of sampler points used to sample the posterior probability space from 1000 to 4000 and find that the posterior probability distribution does not change, demonstrating the robustness of the analysis. As part of our work, we repackage \textit{X-PSI} and its software dependencies into a Docker container. This aids in the portability of the data and the software and streamlines the reproduction of the original analysis. The container is fully documented and contains the scripts for the entire workflow used in our reanalysis. Scripts to reproduce the container are available from a GitHub repository indexed by Zenodo\footnote{\url{https://doi.org/10.5281/zenodo.10564150}}.

This article is organized as follows. First, we describe the original analysis and provide background information on measuring the mass and radius of neutron stars from X-ray data. Then, we describe our effort to reproduce the study and note the computational challenges. Finally, we summarize the lessons we learned when reproducing the Riley~\textit{et al.} analysis and provide guidelines for improvement of the reproducibility of such computationally intensive analyses. 
\section{ANALYSIS OF PSR~J0030+0451}\label{sec:j0030_analysis}

\begin{figure*}[h]
    \centering
\includegraphics[width=\textwidth]{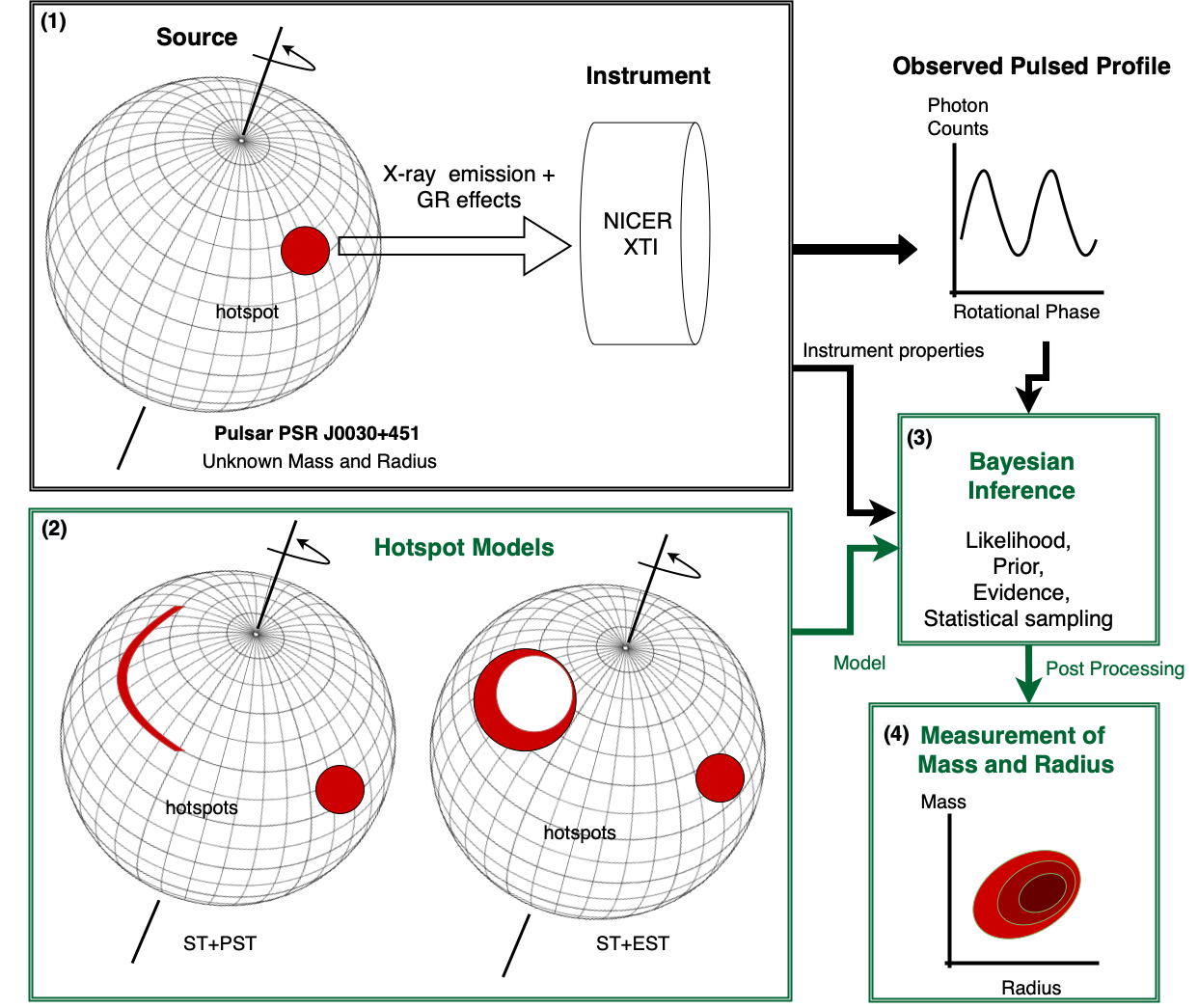}
    \caption{Schematic showing how the mass and radius of PSR J0030+0451 can be measured using X-ray data observed by NICER. The parts of the analysis done by Riley \textit{et al.} using \textit{X-PSI} are shown as green boxes. \textit{X-PSI} uses the data observed by NICER (Box 1 and the observed pulse profile), and the hotspot models that simulate the X-ray emission from the pulsar (Box 2), to perform Bayesian parameter estimation (Box 3) and measure the posterior probabilities for mass and radius of the target pulsar (Box 4). The two examples of hotspot models shown in the figure are `ST+PST' (Single Temperature + Protruding Single Temperature) and `ST+EST' (Single Temperature + Eccentric Single Temperature). }
    \label{fig:overview_fig}
\end{figure*}

Figure~\ref{fig:overview_fig} shows a schematic overview of the analysis by Riley \textit{et al.} including the observation of the X-rays by NICER, modeling of the X-ray emission from the pulsar surface, and estimation of the mass and radius of the pulsar using the observational data and the models. The parts of analysis performed by Riley \textit{et al.} using \textit{X-PSI} are shown as green boxes.

The mass and radius of the neutron star are imprinted on the X-rays emitted by hot spots on the neutron star's surface through the relativistic effect of their propagation through the spacetime curvature induced by the star. The X-ray pulse profile detected by a distant observer encodes the neutron star's compactness, the ratio of the star's mass to its radius. NICER measures X-ray counts as a function of time for a target pulsar, as illustrated in Box 1 of Figure~\ref{fig:overview_fig}. Since the photon count profile of the signal is identical for each rotation of the pulsar, the signal can be phase-folded into a single pulse profile, which gives photon count as a function of the phase of the rotation of the pulsar, as shown in Figure \ref{fig:overview_fig} on the top right. Creating a phase-folded data set is a pre-processing step performed by the NICER instrument team and creates a derived data set used by subsequent analyses. This data set, released using Zenodo, is the starting point for the Riley  \textit{et al.} analyses.

To measure the mass and radius of the star, a model $H$ is created that describes the X-ray emission from the hotspots and uses relativistic ray-tracing of the emitted radiation to predict the pulse profile observed by a distant observer. The parameters of this model are represented by $\vec{\theta}$ and include the mass and radius of the neutron star, the
parameters describing the geometry of the hotspots, the distance to the pulsar, the inclination angle of the axis of rotation to the line of view. In Box 2 of Figure~\ref{fig:overview_fig}, we show the geometry of the hotspots assumed for two models that we use for reproducing the results from the original analysis.

For a given model $H$,  Bayes' theorem is used to infer the posterior probability distribution of the model parameters given a realization of the observed data (Box 3 in Figure~\ref{fig:overview_fig}) according to
\begin{equation}
\label{eqn:bayes} \posterior =
\frac{\likelihood \prior}{\evidence} .
\end{equation}
Here $\prior$ is the probability density of the parameters based on theory, assumptions, or previous observations (the prior); $\likelihood$ is the joint probability distribution as a function of parameters given fixed data (the likelihood); and $\evidence$ is the marginalized likelihood, also called evidence.

Riley \textit{et al.} aims to produce posterior probability measurements of the mass, radius, and other model parameters for a given pulse-profile model and a set of NICER observations of the pulsar J0030+0451. Box 4 in Figure~\ref{fig:overview_fig} shows an example schematic of a two-dimensional marginalized posterior of mass and radius. Each parameter's marginalized posterior probability distribution is obtained using the \texttt{MULTINEST} \cite{ferozMultiNestEfficientRobust2009} implementation of the nested sampling algorithm.

Riley \textit{et al.} explored several hotspot geometry models to determine which model was most favored. However, we only consider the two most likely used models from their analysis and focus on the posterior probabilities for the parameters of these models. To keep the scope of our work reasonable, we neglect models with lower evidence values or higher complexity.

The hotspot geometry used in the most favored model of Riley \textit{et al.} involves two hot regions on the pulsar's surface. The first hotspot is a hot circular disk, whereas the higher temperature in the second hot region lies in the arc-shaped region. This model is named `ST+PST' (Single Temperature + Protruding Single Temperature), shown in Figure \ref{fig:overview_fig}. Using this model, Riley \textit{et al.} found that the mass of the pulsar PSR J0030+0451 is $M = 1.34^{+0.15}_{-0.16}\ M_{\odot}$, (where $1\ M{\odot}$ is the mass of the Sun) and the equatorial radius is $R_{\mathrm{eq}}=12.71^{+1.14}_{-1.19}\ $km. The bounds mentioned here are the $16\%$ and $84\%$ quantiles from the posterior distribution obtained. The compactness $M/R_{\mathrm{eq}}$ is measured to be $0.16^{+0.01}_{-0.01}$. In comparison, Miller \textit{et al.} \cite{Miller:2019cac} measured the mass and radius to be $M = 1.44^{+0.15}_{-0.14}\ M_{\odot}$ and $R_{\mathrm{eq}}=13.02^{+1.24}_{-1.06}\ $km. The second Riley \textit{et al.} model we investigate is `ST+EST' (Single Temperature + Eccentric Single temperature), which differs from ST+PST in that the second hot region is an eccentric annular ring. The model ST+EST gives larger radius and mass ($R_{\mathrm{eq}}=13.89^{+1.14}_{-1.30}\ $km and $M = 1.46^{+0.17}_{-0.18}\ M_{\odot}$, respectively). 

\section{COMPUTATIONAL CONSIDERATIONS}\label{sec:computation}

\begin{figure*}[h]
    \centering
\includegraphics[width=\textwidth]{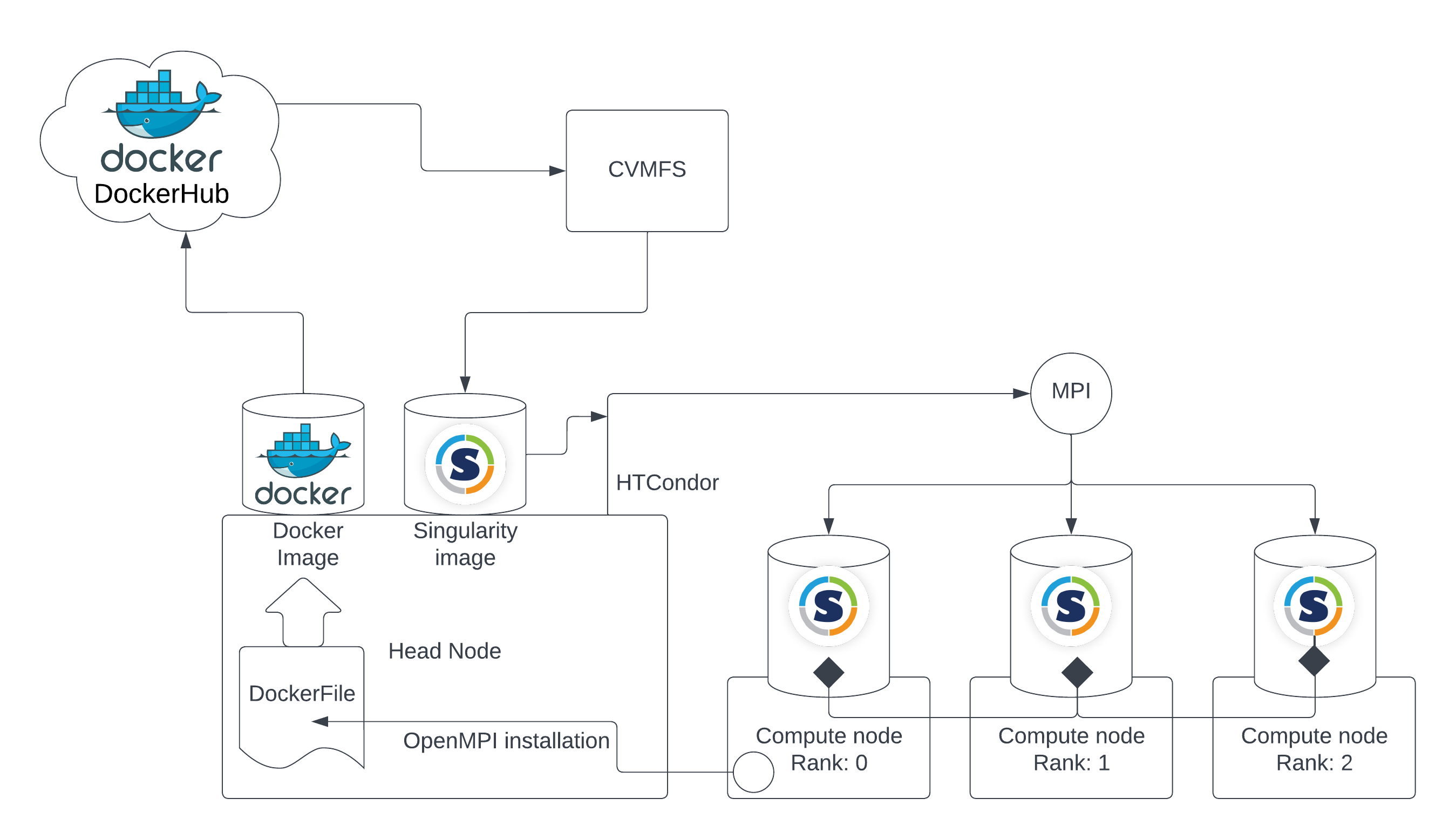}
    \caption{Schematic showing the software components used in our workflow execution. We create a Docker image on the login node on the \texttt{sugwg-condor} cluster, which has the replica of the OpenMPI installation present on any of the compute nodes. This Docker image is then pushed to Docker Hub cloud storage. CVMFS converts the Docker container into a Singularity container and makes it available for use on the cluster. We use the HTCondor job scheduler to deploy the Singularity containers on the compute nodes, thus fulfilling the parallelization requirements of the analysis.}
    \label{fig:software_schematic}
\end{figure*}

The \textit{X-PSI} code used by Riley \textit{et al.} is an open-source code written primarily in Python 2.7, with additional Cython support. As noted in our previous work, Python code presents challenges in reproducibility due to its need for libraries that may not be installed (or may be installed at a different version) on the platform where the code is executed to reproduce an analysis. To address this, Riley \textit{et al.} provided a Python 2.7 Conda environment to install the code and its dependencies. Although this does not isolate the code in the same way as containerization, it facilitates the installation of \textit{X-PSI} and necessary libraries at the correct version.

The documentation provided by Riley \textit{et al.} indicated that they used v0.1 of \textit{X-PSI}  to infer the properties of PSR J0030+0451; this tagged code was made available on GitHub and was straightforward to obtain and install. Riley \textit{et al.} also released a Zenodo repository\footnote{\url{https://doi.org/10.5281/zenodo.5506838}} that contains the phase-folded X-ray data from NICER used as input to the analysis, the configuration files for \textit{X-PSI} v0.1, the submit scripts for the job, the output files of the job, and the files containing the posterior samples for each analysis. This thorough release of data and configuration files makes it possible to reproduce the original analysis, given sufficient computational resources. The repository comes with a \texttt{README.txt} that briefly describes each file and its use. For our analysis, we use the configuration files from the repository, changing the paths to the input data and output files wherever necessary. Following the methods of Riley \textit{et al.} we start our analysis from the phase-folded data and do not attempt to reproduce this from the raw NICER data.

The \textit{X-PSI} analysis is computationally expensive. For example, obtaining marginalized posteriors for the ST+PST Bayesian analysis took 42,453 wall clock hours (see Table 2. from Riley \textit{et al.}). The analysis involves $\mathcal{O}(10^8)$ likelihood evaluations each taking $\mathcal{O}(1)$ second of evaluation time. The likelihood calculation involves simulating hotspots on the star's surface, ray tracing the radiation to include relativistic effects, and creating an instance of the X-ray pulse that a distant observer would detect, making likelihood evaluation the most expensive step. While running \textit{X-PSI} on a single compute node was straightforward, the likelihood evaluations must be executed on multiple compute nodes in parallel to complete the analysis within a reasonable time. To execute the analysis in parallel across multiple compute nodes, \textit{X-PSI} uses the Message Passing Interface (MPI) library.

To distribute the software and input data on each compute node, we used the CERN Virtual Machine File System (CVMFS). This shared filesystem makes scientific software easily available and accessible on HPC clusters. The software stack for interprocess communication uses the \texttt{mpi4py} library~\cite{DALCIN20051108} to create Python bindings to MPI libraries written in C++ and installed as dynamic shared libraries. The object code in these libraries executes the inter-process communication using system calls. Our main reproducibility challenge was to produce a containerized version of \textit{X-PSI} that could execute the analysis using MPI across multiple (possibly heterogeneous) compute nodes.

Following our previous experience, we created a Docker image containing the complete \textit{X-PSI} software stack for execution as a stand-alone image. Using a base Debian Miniconda image\footnote{\url{https://hub.docker.com/r/continuumio/miniconda}}, we installed \textit{X-PSI} and the required dependencies in it using the files provided by Riley~\textit{et al.}. While the Docker container streamlines the installation of \textit{X-PSI} and its dependencies, it presents a problem of running the analysis in parallel on multiple compute nodes as code running each Docker image on a compute cluster needs to communicate with the other images. While this is possible, it is challenging without administrative control of the host machines.

Singularity provides a controlled, containerized environment with the advantage that codes running in the image can access the network capabilities of the host. Unlike Docker's full virtual machine containerization, Singularity creates images that overlay on the host machine. Therefore, if the host's operating system is configured to allow interprocess communication for MPI (as is common in cluster environments), it can be used by code running in the Singularity image. For this to work, the Singularity image must contain the exact version of the \texttt{mpi4py} and the MPI-shared libraries as the host machine. To address this challenge, we used scripts by the Open Science Grid team that converted Docker containers into Singularity images. 

We use HTCondor as the job scheduler and the Syracuse University Gravitational-wave Group (SUGWG) cluster for our analysis. This is a heterogeneous combination of Intel® Xeon® Gold 6248R @3.00GHz, E5-2660v2 @2.20GHz, E5-2698v3 @2.30GHz, X5650 @2.67GHz, X5550 @2.67GHz, E5-2620 0 @2.00GHz, and AMD EPYC 7702P, EPYC 7543 processors). The cluster uses a CentOS operating system configured to allow codes to use OpenMPI~\cite{10.1007/978-3-540-30218-619} implementation of MPI. 

Since the Docker container (and hence the derived Singularity image) used to host \textit{X-PSI} has a Debian operating system, we made a copy of the OpenMPI shared libraries that are installed on the SUGWG compute nodes, deployed it in the container, and configured the runtime linker so that the Python interpreter inside the container could access these libraries. This ensures that the OpenMPI within the container has paths and configurations identical to the host compute nodes. The analysis can then be launched using the HTCondor job scheduler. HTCondor uses \texttt{mpirun} to execute \textit{X-PSI} from inside the Singularity container across multiple compute nodes. Figure \ref{fig:software_schematic} shows the software setup schematic.

\section{REPRODUCING THE J0030 RESULT}
\label{sec:results}

\begin{table*}[]
    \begin{tabular}{@{}m{.8cm}m{3.6cm}@{\hspace{.3cm}}>{\raggedright\arraybackslash}m{1.5cm}@{\hspace{.3cm}}>{\raggedright\arraybackslash}m{1.5cm}@{\hspace{.3cm}}>{\raggedright\arraybackslash}m{2cm}@{\hspace{.3cm}}>{\raggedright\arraybackslash}m{2cm}@{\hspace{.3cm}}>{\raggedright\arraybackslash}m{2cm}@{}}
    \toprule[2pt]
    \multicolumn{2}{@{}l}{\textbf{Model}} &
    \textbf{Results in Riley \textit{et al}.} &
    \textbf{Data from Zenodo release} &
    \textbf{Reproducibility analysis} & 
    \textbf{Analysis using broader radius priors} &
    \textbf{Analysis using 4000 sampler live points} 
    \\
    
     \midrule[1pt]
     \multirow{11}{*}{ST+PST} & $\ln{Z}$ & -36368.28 & -36366.65 & -36365.52 & -36365.24 & -36364.39 \\[3pt]
        & Mass [$M_{\odot}$] & $1.34^{+0.15}_{-0.16}$ & $1.34^{+0.15}_{-0.15}$ & $1.34^{+0.16}_{-0.15}$ & $1.36^{+0.16}_{-0.16}$ & $1.35^{+0.16}_{-0.16}$ \\[5pt]
        & Equatorial Radius [km] & $12.71^{+1.14}_{-1.19}$ & $12.7^{+1.1}_{-1.2}$ & $12.8^{+1.2}_{-1.2}$ & $12.9^{+1.2}_{-1.2}$ & $12.9^{+1.3}_{-1.2}$ \\
        \cmidrule(l){2-7} 
        & Likelihood evaluations & 78,343,018 & 78,343,018 & 157,814,515 & 139,593,698 & 589,513,174 \\
        & Nested Replacements & 57,972 & 57,972 & 56,896 & 56,596 & 225,856 \\
        & Weighted Posterior Samples (?) & 20,177 & 12,242 & 11,896 & 11,749 & 46,488 \\
        \cmidrule(l){2-7} 
        & CPU hours & 42,453 & 42,453 & 48,384 & 55,296 & 179,712 \\
        & Number of cores & 960 & 960 & 288 & 384 & 288 \\
     \midrule[1pt]
     \multirow{10}{*}{ST+EST} & $\ln{Z}$ & -36367.81 & -36366.17 & -36366.14 & -36366.16 & - \\[3pt]
        & Mass [$M_{\odot}$] & $1.46^{+0.17}_{-0.18}$ & $1.46^{+0.17}_{-0.18}$ & $1.46^{+0.17}_{-0.17}$ & $1.47^{+0.19}_{-0.19}$ & - \\[5pt]
        & Equatorial Radius [km] & $13.89^{+1.14}_{-1.30}$ & $13.9^{+1.1}_{-1.3}$ & $13.8^{+1.2}_{-1.2}$ & $14^{+1.4}_{-1.4}$ & - \\
         \cmidrule(l){2-7}    
        & Likelihood evaluations & 88,965,106 & 88,965,106 & 89,850,127 & 143,920,078 & - \\
        & Nested Replacements & 53,149 & 53,149 & 53,098 & 52,358 & - \\
        & Weighted Posterior Samples (?) & 20,177 & 12,242 & 10,944 & 10,828 & - \\
        \cmidrule(l){2-7} 
        & CPU hours & 61,210 &  61,210 & 48,384 & 55,296 & - \\
        & Number of cores & 960 & 960 & 288 & 480 & - \\ 
        \bottomrule[2pt]
    \end{tabular}
    \vspace{2mm}
    \caption{Summary of job statistics for the original analysis published by Riley \textit{et al.} in the Zenodo repository, and the reanalyses. We show the key results in the first section of the models ST+PST and ST+EST for the Bayesian evidence ($\ln{Z}$) obtained from the analysis, and the measurement of mass and the equatorial radius of the target pulsar. The results in Riley \textit{et al.} are obtained after post-processing the data they obtained from the job. The latter has been publicly released in the Zenodo repository, so the computational details (CPU hours, number of cores) for these two columns will be the same. In contrast, the evidence values and measurements of mass and radius will differ. }
    \label{tab:job_statistics}
\end{table*}{}

We reproduce the original analyses using the ST+PST and ST+EST models using the above software setup. Additionally, we submit jobs using the ST+PST and ST+EST models with broader radius priors than the original analyses. All analyses were run with a sampling efficiency of $0.3$ and an evidence tolerance of $0.1$, as used by Riley \textit{et al.}. Table \ref{tab:job_statistics} shows the job statistics and summary of results obtained for the analysis in Riley \textit{et al.} and for our work in this paper. We also show the information provided in the Zenodo release by Riley \textit{et al.}. The results from the data in the Zenodo repository and those in the publication are different because Riley \textit{et al.} post-processed the posterior samples (present in the Zenodo repository) to re-evaluate the evidence. We do not repeat this post-processing step since we do not have access to a working post-processing script used by the original study's authors.

\begin{figure*}[t]
\begin{centering}
\hspace{-0.5cm}
\includegraphics[width=\textwidth]{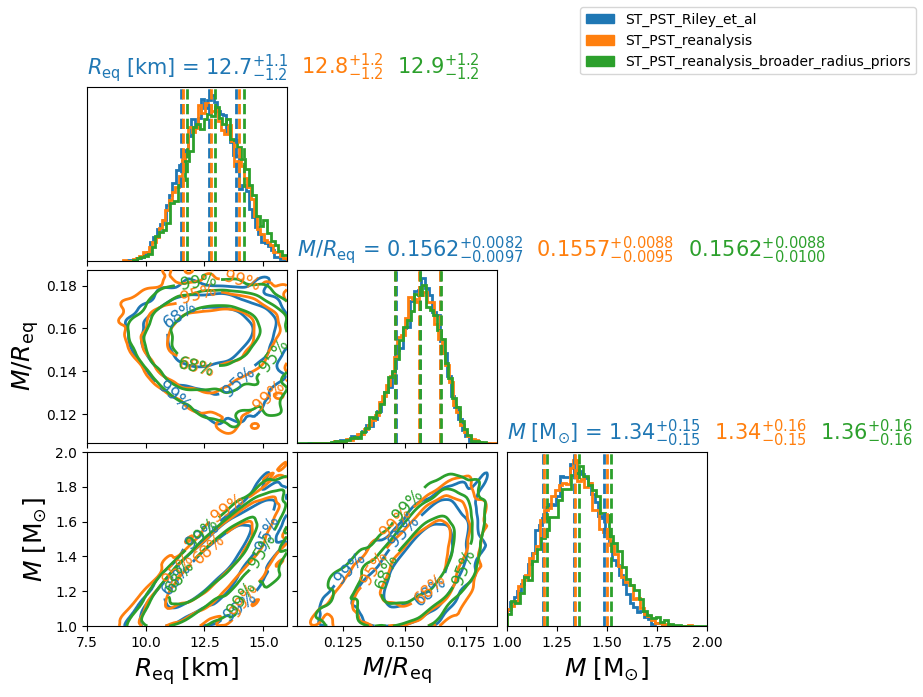}
\end{centering}
\caption{Comparison of posterior probability distributions for mass, radius, and compactness of J0030 obtained by Riley \textit{et al.} (blue), reproducing the analysis (orange), and for the analysis with the broader radius priors (green), using the hotspot model ST+PST. The corner plot shows the three parameters' one and two-dimensional marginal posteriors. The priors used for the re-analysis are the same as in the original analysis.}
\label{fig:posterior_j0030_ST_PST}
\end{figure*}

\begin{figure*}[t]
\begin{centering}
\includegraphics[width=\textwidth]{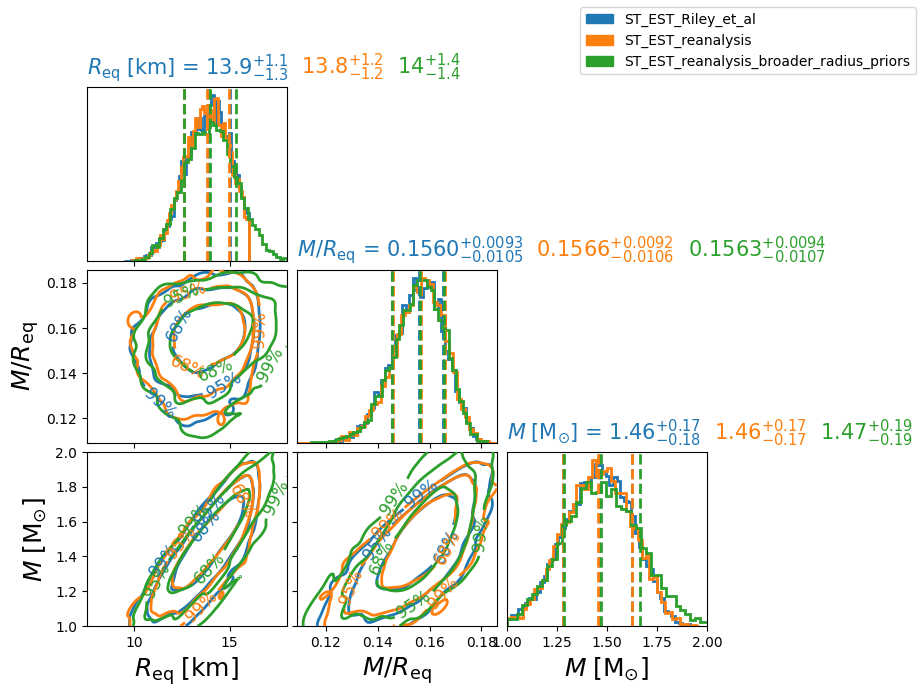}
\end{centering}
\caption{Comparison of posterior probability distributions for mass, radius, and compactness of J0030 obtained by Riley \textit{et al.} (blue), reproducing the analysis (orange), and for the analysis with the broader radius priors (green), using the hotspot model ST+EST. The corner plot shows the three parameters' one and two-dimensional marginal posteriors. The priors used for the re-analysis are the same as in the original analysis.}
\label{fig:posterior_j0030_ST_EST}
\end{figure*}

Figure \ref{fig:posterior_j0030_ST_PST} shows the posteriors for the model ST+PST from the Zenodo repository (results used by Riley \textit{et al.}) in blue, from the reproducibility analysis in orange, and for the analysis with the broader radius priors in green. Figure \ref{fig:posterior_j0030_ST_EST} shows when the model ST+EST was used. The figures show one and two-dimensional marginal posteriors for the mass $M$ (in solar masses), the equatorial radius $R_{eq}$ (in km), and the compactness $M/R_{eq}$ (in solar mass/km) of the pulsar. The original and reanalysis analysis involved 19 parameters (for ST+PST and ST+EST models). For brevity, we show the posteriors for only three parameters, mass, radius, and compactness.

We reproduced the result from the original analyses. We get the exact measurement for the three quantities of interest with the same $68$ percentile confidence interval. The minor differences in the posteriors are statistical, and one expects this order of fluctuation after each repetition of the analysis. Since Nested sampling is a Monte Carlo technique, one cannot obtain identical reproduction of the positions of the sample points. The samples exploring the posterior space accumulate around the region with high probability. There would be some fluctuation at the periphery of this distribution, which is reflected in the deviations in the $99\%$ contour lines, where the sampler population density is sparse. The inner $68\%$ and $95\%$ contour lines in the posterior from the reanalysis show much less deviation from the posterior of the original analysis. 

Although the value for evidence we get for the reanalysis is close to the value reported in the original paper, the difference is larger than would be expected from the evidence tolerance provided to MultiNest. Although statistical fluctuations are expected in the evidence, the values obtained illustrate the challenges in comparing evidences between models even when the physical posterior parameters are in good agreement.

Post-processing and plotting the output of the Bayesian analysis proved to be an obstacle to the reproducibility of the original results. Although the Zenodo repository had all the configuration files and submitted scripts to start the analysis, the post-processing and plotting scripts were absent. These scripts are necessary to produce a figure that is identical to the figure that was published. The documentation of \textit{X-PSI} describes the post-processing module of the software. However, the documentation describes v0.5 of \textit{X-PSI}, which is backward incompatible with v0.1 used in the original analysis. We found that the post-processing module of \textit{X-PSI} failed to process the output files from \texttt{MULTINEST}. Instead, we used the post-processing modules and scripts from PyCBC Inference \cite{Biwer:2018osg}--a Python toolkit for Bayesian analysis of gravitational-wave signals--to plot all the posteriors. We converted the \texttt{.dat} files produced by the \texttt{MULTINEST} sampler into PyCBC-readable \texttt{.hdf} files and used the \texttt{pycbc\_plot\_posterior} script on these files. This emphasizes the importance of releasing the set of all the executables used in the original analysis, including the post-processing and the plotting scripts. 

In addition to reproducing the original analyses, we also explored the effects of using broader prior bounds for the radius. The original analysis put the upper bound at 16 km for the neutron star's radius. We changed it to 25 km and found that the posteriors are unaffected. This test is useful for the model ST+EST, where the posteriors are cut off at the upper bound of the prior. We aimed to check if the posteriors were affected if the prior base was increased. Since the posteriors did not change significantly, we conclude that the data observed by the NICER instrument is informative and that our choice of priors does not heavily influence the Bayesian analysis. The posteriors for the analyses with broader radius priors for models ST+PST and ST+EST are shown in green in Figures \ref{fig:posterior_j0030_ST_PST} and \ref{fig:posterior_j0030_ST_EST}, respectively. 

For the model ST+PST, we also perform an analysis with an increased number of live points used by the sampler to sample the posterior probability distribution. The original analysis used 1000 live points, and we increased it to 4000 live points to check the robustness of the result to the sampler configuration. The posteriors do not significantly change when the number of live points for the sampler increases.
\section{LESSONS LEARNED}
\label{sec:lessons_learned}

We compiled a list of challenges we encountered while reproducing the analysis and noted the lessons learned. We discuss the guidelines to make computational analysis, such as that done by Riley \textit{et al.}, reproducible. Table \ref{tab:availability} lists whether the data, software, and documentation components were available, incomplete, or unavailable before our reproducibility study. 

\begin{table*}[]
    \centering
    \normalsize
    \begin{tabular}{llll}
    \toprule[2pt]
    \multicolumn{3}{@{}l}{\textbf{Data}} \\
    \midrule[.5pt]
     & Raw input data        & Unavailable  & \\     
     & Processed input data  & Available & \url{https://doi.org/10.5281/zenodo.5506838} \\ 
     & Output data & Available & \url{https://doi.org/10.5281/zenodo.5506838} \\
     \midrule[1.5pt]
    \multicolumn{3}{@{}l}{\textbf{Software}}                                                           \\ \midrule[.5pt]
      & Code  & v0.1 Available~\cite{xpsi}  &  \url{https://github.com/xpsi-group} \\              
      & Documentation & Available for v0.5  & \url{https://xpsi-group.github.io/xpsi/index.html} \\
     & Software dependencies & Available &  \\
     & Configuration files   & Available &  \\ 
     & Post-processing scripts & Unavailable & \\ 
     \midrule[1.5pt]
   
    \end{tabular}%
        \vspace{2mm}
    \caption{Availability of data, scripts, code, and documentation before our reproducibility study.}
    \label{tab:availability}
\end{table*}

{\bf Input Data Availability.} The raw data for the NICER observation of PSR~J0030+451 was not made available by Riley \textit{et al.} through their Zenodo repository. However, processed data was included in the Zenodo release and had accompanying documentation.

{\bf Software Availability.}
Riley \textit{et al.} use \textit{X-PSI} v0.1 to analyze PSR~J0030+451 data. The code is open source and publicly available on GitHub.

{\bf Software Documentation.} \textit{X-PSI} comes with extensive, publicly-accessible documentation. Since the framework of the code is modular, the documentation goes over each module in depth, explaining the physics associated with it and providing examples. However, the code and its documentation have evolved significantly since they were used for the original analysis. The documentation during our reproducibility effort relates to \textit{X-PSI} v0.5, whereas the original analysis used \textit{X-PSI} v0.1. 

{\bf Software Installation and Dependencies.} The instructions for installation of \textit{X-PSI} include information about all the software dependencies. They also provide \texttt{.yml} files that can be used to create a virtual environment with the basic dependencies resolved. The installation manual has clear instructions for installing the sampler and the parallelization software. 

{\bf Configuration files.} The Zenodo repository has all the configuration files used by \textit{X-PSI} to generate the hotspot models. The availability of the configuration files was crucial to successfully reproducing the results. The configuration files shared in the Zenodo repository streamline the setup of the jobs. Combined with the documentation, modifying the original analysis and changing the sampler configuration and the prior bounds for radius was easy.

{\bf Computational Resources} The original analysis of Riley \textit{et al.} used the Dutch national SURFsara supercomputer Cartesius. As is common in attempts to reproduce analyses, we did not have access to these computational resources or the original environment used by Riley \textit{et al.}. To execute \textit{X-PSI} on the large-scale resources available to us, we had to adapt the \textit{X-PSI} deployment to fit a different scheduler and create an overlay container that could run on this cluster. By demonstrating that this is possible, we show that overcoming the barrier to reproducibility presented by the lack of access to computing resources is possible.

{\bf Post-processing Scripts.} The unavailability of the post-processing scripts to analyze the output data from the analysis and plot the posteriors made it unfeasible for us to generate the same plot as the one present in the publication of the original results. We had to use PyCBC software to plot the posterior distributions. Post-processing scripts are a crucial part of reproducibility in the software workflow. The scripts to process the raw data are also absent. 

{\bf Output Data Availability.} The Zenodo repository included the posterior output files for all the analysis jobs performed by the original study's authors. The output files included all the files generated by the sampler \texttt{MULTINEST}, including the posterior file and the history of all the sample points throughout the analysis. It also includes the output of the jobs --- including the job scheduler logs and error messages generated by \textit{X-PSI} during the analysis. 

\section{CONCLUSIONS}
\label{sec:conclusions}

Conducting reproducible research is an essential step towards open science. In this article, we described the procedure and challenges involved in reproducing the measurement of mass and radius of the pulsar PSR~J0030+451 from the X-ray data observed by NICER. 

Given the release of the Zenodo repository containing the data and the configuration scripts used for the original analysis, we were able to reproduce the analysis by Riley \textit{et al.} to measure the mass and the radius of PSR~J0030+451. The post-processing scripts plot the posteriors using the output file produced by \textit{X-PSI} is absent. We could not use the code and its documentation to plot the posteriors as shown in the original publication. Instead, after converting the output file to an \texttt{.hdf} file, we used the post-processing module of PyCBC to plot the posteriors. This highlights the importance of releasing the entire set of scripts, from data processing to post-processing of the analysis output, to be released in a containerized format to reproduce the analysis. 

Apart from reproducing the measurement, we changed the prior probabilities of the radius from the original analysis, increasing the upper bound from 16 km to 25 km. Despite the broader range of possible radii from the prior, we get the exact posterior distribution as the original analysis. We also increased the number of points used by the sampler from 1000 to 4000 and found no significant change in the posterior probability distribution.

Our work also shows that it is possible to reproduce analyses that require large-scale computational resources without access to the original hardware. This is significant, as access to resources is often a major barrier to reproducibility. Scientists wishing to reproduce findings might not have allocations on the original resources, or the original resource may have been decommissioned. Using the Singularity overlay container shows that executing MPI code across a heterogeneous cluster that uses a different operating system than the original hardware is possible.

To aid future researchers who want to reproduce the analysis of PSR J0030+451 data, the Docker image created for our analysis is publicly available\footnote{\url{https://hub.docker.com/r/chaitanyaafle/nicer}}. The specific tag of the container's image used for the reproducibility analyses is `8d3b23d`. The Dockerfile is also available publicly on the GitHub repository accompanying this article \footnote{\url{https://github.com/sugwg/nicer-reproducibility-J0030}}. We provide the post-processing script and the PyCBC installation required to produce the posterior corner plots.

\section{ACKNOWLEDGEMENT}
We thank Anna Watts for the helpful discussions. The U.S. National Science Foundation supported this work under Grants 2207264, 2041977, 2041901, 2028923, 2028930, 1841399, and 1941443. 
The work of I.T. was supported by the U.S. Department of Energy, Office of Science, Office of Nuclear Physics, under contract No.~DE-AC52-06NA25396 and by the Laboratory Directed Research and Development program of Los Alamos National Laboratory under project numbers 20220541ECR and 20230315ER.
Syracuse University provided computational resources. 

\bibliographystyle{IEEEtran}
\bibliography{references}

\begin{IEEEbiography}
{Chaitanya Afle}{\,}is a Ph.D. candidate in Physics at Syracuse University, NY. His research interests include gravitational-wave astronomy and astrophysics, data analysis, and multimessenger astronomy.
\end{IEEEbiography}

\begin{IEEEbiography}
{Patrick R. Miles}{\,}obtained his M.S. in Physics from Syracuse University in 2021.
\end{IEEEbiography}

\begin{IEEEbiography}
{Silvina Caíno-Lores}
is a Research Assistant Professor at the University of Tennessee, Knoxville. She obtained her Ph.D. in Computer Science and Technology from Carlos III University of Madrid (Spain) in 2019. Her research interests include cloud computing, in-memory computing and storage, HPC scientific simulations, and data-centric paradigms. 
\end{IEEEbiography}

\begin{IEEEbiography}
{Collin D. Capano}
is a senior scientist at the Max Planck Institute for Gravitational Physics and High Performance Computing Facilitator at the University of Massachusetts Dartmouth. He obtained his PhD from Syracuse University and his research focuses on the analysis of gravitational waves from black holes and neutron stars.
\end{IEEEbiography}

\begin{IEEEbiography}
{Ingo Tews}
is a staff scientist at Los Alamos National Laboratory. He obtained his Ph.D. from Technische Universität Darmstadt and is interested in fundamental interactions between neutrons and protons in atomic nuclei and neutron stars.
\end{IEEEbiography}

\begin{IEEEbiography}{Karan Vahi}{\,}is a Senior Computer Scientist at USC Information Sciences Institute. Vahi received a M.S in Computer Science in 2003 from the University of Southern California. 
His research interests include scientific workflows and distributed computing systems. Contact him at vahi@isi.edu.
\end{IEEEbiography}

\begin{IEEEbiography}{Ewa Deelman}{\,} received her Ph.D. in Computer Science from the Rensselaer Polytechnic Institute. She is a Research Director at USC/ISI and a Research Professor at the USC Computer Science Department and leads the design and development of the Pegasus Workflow Management software.  Her research explores the interplay between automation and the management of scientific workflows that include resource provisioning and data management. 
\end{IEEEbiography}

\begin{IEEEbiography}{Michela Taufer}{\,} holds the Dongarra Professorship in High-Performance Computing within the Department of Electrical Engineering and Computer Science at the University of Tennessee, Knoxville. Dr. Taufer received her Ph.D. in computer science from the Swiss Federal Institute of Technology (ETH) in 2002. 
Her interdisciplinary research is at the intersection of computational sciences, high-permanence computing, and data analytics. 
\end{IEEEbiography}

\begin{IEEEbiography}{Duncan A. Brown}{\,} is the Charles Brightman Professor of Physics at Syracuse University. Dr.~Brown received a Ph.D. degree in physics from the University of Wisconsin-Milwaukee in 2004. He was a member of the LIGO Scientific Collaboration from 1999 to 2018 and is a fellow of the American Physical Society. 
His research is in gravitational-wave astronomy, astrophysics, and large-scale scientific workflows. Contact him at dabrown@syr.edu.
\end{IEEEbiography}

\end{document}